\documentclass[twocolumn, secnumarabic, amssymb, amsmath, nofootinbib,tightenlines, showpacs, nobibnotes, aps, prl]{revtex4}
\usepackage{amsfonts}

\usepackage{bm}%
\usepackage[colorlinks=true,linkcolor=blue]{hyperref}%
\usepackage{graphicx}
\usepackage{dcolumn}

\begin{document}

\title{Phase Diagram of a Spin-Orbit Coupled Fermi Gases in a Bilayer Optical Lattice}

\author{Xiaosen Yang$^{1}$}
\author{Beibing Huang$^{2}$}
\author{Hai-Qing Lin$^{1}$}
\affiliation{$^{1}$ Beijing Computational Science Research Center,
Beijing, 100084, \textbf{P. R. China}}
\affiliation{$^{2}$ Department of Experiment Teaching, Yancheng Institute of Technology,
Yancheng, 224051, \textbf{P. R. China}}
\date{\today}

\begin{abstract}
We investigate the stability of helical superfluid phase in a spin-orbit coupled Fermi gas loaded in a bilayer optical lattice. The phase diagram of the system is constructed in the mean field framework. We investigate the topological properties of the superfluid phases by a nontrivial application of the Fermi surface topological invariant to our time-reversal invariant system with degeneracies on the Fermi surface. We find that there is a first-order phase boundary in the phase diagram of half filling case and the superfluid phases are all topological trivial. The superfluid phase is topological nontrivial when the filling fraction deviates from the half filling. In the topological nontrivial superfluid phase, a full pairing gap exists in the bulk and gapless helical Majorana edge states exist at the boundary.

\end{abstract}


\maketitle

\section{Introduction}
Recently, the topological properties have been investigated extensively in condensed matter systems such as topological insulators (TIs) \cite{M. Z. Hasan,X.l.Qi2011rmp, X.l.Qi,Zhong Wang, ZWang2012a},  topological  superfluids (TSFs)/superconductors (TSCs) \cite{ N. Read, M. Sato, A. Kubasiak, schnyder, Tewari,ZWang2012b,ZWang2012c}. TSFs/TSCs have a full pairing gap in the bulk and gapless Majorana edge states at the boundary. These new quantum phases are described by topological order \cite{X. G. Wen} instead of the traditional Landau symmetry breaking theory. The gapless Majorana edge states are protected by the symmetry of bulk for the bulk-edge correspondence. For the application, the gapless Majorana edge states have significant application in topological quantum computation and attract considerable attention \cite{XLQi2009, M. Gong2}. In two dimensions(2D), there are two classes of TSFs/TSCs, namely the chiral superfluids/superconductors and helical superfluids/supconductors. Chiral superfluids/supconductors break the time-reversal symmetry and have chiral Majorana edge states at the boundary \cite{N. Read, M. Sato, Tewari}. Whereas the helical superfluids/supconductors are time-reversal invariant (TRI) and have gapless helical Majorana edge states at the boundary \cite{XLQi2010, shusadeng2012, Nakosai}. They are classified by an integer and a $Z_{2}$ invariant respectively \cite{schnyder}. The helical superfluid phase can be stabilized in a two-band spin-orbit coupled system\cite{shusadeng2012} or a bilayer system\cite{Nakosai} with SOC and repulsive interlayer interaction.

In ultracold fermonic system, TIs/TSFs are proposed to be realized by utilizing spin-orbit coupling (SOC). Most recently, the effective SOC has been successfully realized both in ultracold bosonic \cite{Y. J. Lin1, Y. J. Lin2} and fermonic \cite{P Wang, Cheuk} systems and has generated a considerable amount of theoretical interests \cite{S. L. Zhu, Vyasanakere, Z. Q. Yu, H. Hu, Wei yi, Wei yi2, M. gong, Iskin, Salasnich, L. Han, J. Liu, L. He,F. Wu}.  With remarkable tunability and clean environment \cite{Leggett, Giorgini, Bloch}, ultracold Fermi gases provide an ideal platform for investigating the interesting topological properties. For a spin-polarized Fermi gas, the chiral superfluid phases can be stabilized in the presence of spin-orbit coupling and exhibit many new interesting physics \cite{M. Sato,A. Kubasiak, J zhou, X yang1, X yang2}. However, the study of the helical superfluid phase are destitute in ultracold fermonic system. In addition, a bilayer fermionic system exhibits many new interesting phenomena \cite{DWWang2007, Pikovski, Zinner} compared to the singlelayer case. With a SOC, the bilayer fermionic system can stabilize the helical superfluid phase without breaking the time-reversal symmetry.

In this paper, we investigate a spin-orbit coupled Fermi gas loaded in a bilayer optical lattice by using a BCS-type mean field theory at zero temperature. The interlayer hopping is included instead of the interlayer interaction. The ground state is self-consistently determined by minimizing the thermodynamic potential. Due to the degeneracies on the Fermi surfaces, we investigate the topological properties of the superfluid phases by a nontrivial application of the Fermi surface topological invariant (FSTI). Our main results are the followings. First, the superfluid phases are all topological trivial in the phase diagram of half filling case while there is a first-order phase boundary. Second, the superfluid phase is topological nontrivial when the filling fraction deviate from the half filling. Without breaking the time-reversal symmetry (TRS), the topological nontrivial superfluid phase is helical superfluid phase. This state has a full pairing gap in the bulk and gapless helical Majorana edge states counterpropagating at the boundary. In the phase diagram of filling fraction, there are two helical superfluid phases which are separated by a gapless boundary. There is no difference between the two helical superfluid phases in topology. Lastly, the effect of the trapping potential is discussed under local density approximation. Our work provides a possible route to realize the helical superfluid phase in ultracold fermionic system in the future.

\section{Formalism of the system}
We consider a system of bilayer fermi gas of isotropic spin-orbit coupling. In the mean field framework, the system's Hamiltonian can be given as the following:
\begin{eqnarray}
H=H_{0}+H_{SO}+H_{s}, \label{1}
\end{eqnarray}
where $H_{0}$ is kinetic term, $H_{SO}$ is isotropic SOC term and $H_{s}$ is s-wave pairing term. The following are their detail expression
\begin{eqnarray}
&&H_{0}= \sum_{<i,j>}  \psi_{i}^{\dag} (t - \mu \delta_{i,j} + t_{h} \delta_{i,j} \tau_{x} ) \psi_{j},\\
&&H_{SO}= - \sum_{i,\hat{{\bf  e}}=\hat{{\bf x}},\hat{{\bf y}}} [ \lambda \psi_{i}^{\dag} (i \hat{{\bf \sigma}} \times \hat{{\bf e}})_{z} \tau_{z}  \psi_{i+\hat{{\bf e}}} + h.c.],\\
&&H_{s}= -\sum_{i} ( \Delta a_{i \uparrow}^{\dag} a_{i \downarrow}^{\dag}  - \Delta b_{i \uparrow}^{\dag} b_{i \downarrow}^{\dag} + h.c.).
\end{eqnarray}
Here,$\psi_{i}=(a_{i,\uparrow},a_{i,\downarrow}, b_{i,\uparrow},b_{i,\downarrow})^{T}$ with $a ( a ^{\dag})$ and $b (b ^{\dag})$ denoting the fermion annihilation (creation) operators for $A $ and $B$ layers, respectively. The Pauli matrices ${\bf \sigma}$ act on the spin and ${\bf \tau} $ acts on the layer, $t_{h}$ is the strength of interlayer hopping. In the above forms, we have assumed that there is a $\pi$ phase shift between the SOCs of the two layers, $\lambda$ is the strength of the SOC. We have also assumed $\Delta_{A}= -\Delta_{B}= \Delta$ \cite{gapphase} with $\Delta_{A}=-U <a_{i\downarrow} a_{i,\uparrow}> $ and $\Delta_{B}= -U <b_{i \downarrow} b_{i \uparrow}>$, where $U$ is the strength of the attractive on-site interaction. We only consider the intralayer on-site interaction and the interlayer hopping for simplicity.

Therefore, by introducing $\psi^{\dag}_{{\bf k}} = (a^{\dag}_{{\bf k}}, b^{\dag}_{{\bf k}}, a_{-{\bf k}}, b_{-{\bf k}})$ with Fourier transformations $a_{i}=\sum_{{\bf k}} \exp( i {\bf k} \cdot {\bf r_{i}}) a_{{\bf k}}/\sqrt{N}$ and $b_{i}=\sum_{{\bf k}} \exp( i {\bf k} \cdot {\bf r_{i}}) b_{{\bf k}}/\sqrt{N}$ ($N$ is the number of the sites of the lattices), the Hamiltonian is given by
\begin{eqnarray}
H= \frac{1}{2}\sum_{{\bf k}} \psi^{\dag}_{{\bf k}}\mathcal{H}({\bf k})\psi_{{\bf k}} + \sum_{{\bf k}} 2 \xi_{{\bf k}},\label{2}
\end{eqnarray}
with
\begin{eqnarray}
&&\mathcal{H}({\bf k})=\nonumber\\
&&\left(
  \begin{array}{cccc}
    \xi_{{\bf k}}+ {\bf g_{k}}\cdot{\bf \sigma} & t_{h} & i \Delta \sigma_{y}  &  0  \\
    t_{h} & \xi_{{\bf k}} - {\bf g_{k}}\cdot{\bf \sigma}   & 0 & -i \Delta \sigma_{y}\\
    -i \Delta \sigma_{y} & 0 & -\xi_{{\bf k}}+ {\bf g_{k}} \cdot {\bf \sigma}^{T}  & -t_{h} \\
    0 & i \Delta \sigma_{y} &-t_{h} &  -\xi_{{\bf k}} - {\bf g_{k}}\cdot{\bf \sigma}^{T}  \\
  \end{array}
\right).
\end{eqnarray}
where, $\xi_{{\bf k}}=-2t(\cos k_{x}+\cos k_{y})-\mu$ and ${\bf g_{k}} = 2 \lambda(\sin k_{y}, -\sin k_{x})$. Here,the Hamiltonian of two dimension Bogoliubov-de Gennes (BdG) possess the partical-hole symmetry (PHS) and TRS. Therefore, the classification of the Hamiltonian is $Z_{2}$ class.
Diagonalizing the Hamiltonian, the excitation spectrum $E_{\pm}({\bf k})$ of the quasiparticles are
\begin{eqnarray}
E_{\pm}({\bf k}) = \sqrt{\Delta^{2} + \xi_{{\bf k}}^{2} + |{\bf g_k}|^{2} + t_{h}^{2} \pm 2 E_{0}}.\label{3}
\end{eqnarray}
with $E_{0} = \sqrt{t_{h}^{2} (\Delta ^{2} + \xi_{{\bf k}}^{2}) + \xi_{{\bf k}}^{2} |{\bf g_{k}}|^{2}}$. The topological phase transition only occurs at the closing points of the gap, that is to say, the topological invariants may change only when the gap closes. In the presence of pairing gap, the excitation spectrum is gapless only when $|{\bf g_{k}}|=0$ and $t_{h} =\sqrt{\xi_{{\bf k}}^{2} + \Delta^{2}}$ which can be simplified as $t_{h} =\sqrt{\xi_{{\bf k_{c}}}^{2} + \Delta^{2}}$ with ${\bf k_{c}} \in \{ (0,0), [(0, \pi),(\pi, 0)],(\pi, \pi)\}$.

The thermodynamic potential is $\Omega = -\mathrm{Tr} \ln [e ^{- \beta H}]$ with $\beta^{-1}=k_{B}T$. At zero temperature, the thermodynamic potential is
\begin{eqnarray}
\Omega=\frac{2|\Delta|^{2}}{U} + \sum_{{\bf k},\nu} (\xi_{{\bf k}}-E_{\nu}({\bf k})).\label{4}
\end{eqnarray}

The pairing gap and the chemical potential should be determined  self-consistently by minimizing the thermodynamic potential. The gap and number equations are given as
\begin{eqnarray}
&&\frac{1}{U}= \sum_{{\bf k},\nu=\pm} \left[\frac{1}{4 E_{\nu}({\bf k})}\left(1 + \nu \frac{t_{h}^{2}}{E_{0}}\right) \right],\\
&&n= \sum_{{\bf k},\nu=\pm } \frac{1}{2}\left[ 1 - \frac{\xi_{{\bf k}}}{E_{\nu}({\bf k})}\left(1+\nu \frac{t_{h}^{2} + |{\bf g_{k}}|^{2}}{E_{0}}\right) \right].
\end{eqnarray}
By self-consistently solving above equations, we have found the phase diagram of system. The topological invariant properties will be discussed in the next section.

\section{Topological properties}
Rigorous topological invariants of interacting insulators and superfluids/superconductors defined in terms of 'topological Hamiltonian',  which is the inverse of zero frequency Green's function $h_{t} = - G^{-1}(0,{\bf k})$ , have been proposed in Ref.\cite{ZWang2012a,ZWang2012b,ZWang2012c} recently. In the mean-field approximation and weakly pairing limit, we can also use the FSTI given by  Qi. et al in Ref.\cite{XLQi2010}, which takes  the following form
\begin{eqnarray}
N_{2D}= \prod_{i}[{\rm sgn}(\delta_{i})]^{m_{i}},
\end{eqnarray}
where $m_{i}$ is the number of TRI points enclosed by the $i$th Fermi surface, ${\rm sgn}(\delta_{i})$ is the sign of pairing gap on the $i$th Fermi surface of single particle Hamiltonian $h_{{\bf k}}$ with
$
\delta_{i,{\bf k}}= <i,{\bf k}| \mathcal{T} {\bf \Delta}^{\dag} |i,{\bf k}>
$,
where $|i, {\bf k}>$ are the eigenvectors of $h_{{\bf k}}$ and $ {\bf \Delta} = i \Delta \sigma_{y} \tau_{z}$. The 2D TRI superfluid phase is topological trivial for $N_{2D}=1$ and nontrivial for $N_{2D}=-1$.
The FSTI shows that the topological properties of a superfluid phase is completely determined by the Fermi surface properties in weakly pairing limit. For our system, the pairing gap can be larger than the strength of the hopping($t$), thus the FSTI is not well defined in these regions. Even though the FSTI can only be applied to the narrow superfluid region near the boundaries of the normal and superfluid phases, the information is adequate for determining the topological invariants of the superfluid phases in the phase diagrams because we expect that topological properties should be the same for the same phase.

The eigenvalues of the single particle Hamiltonian are
$
E_{i}({\bf k}) = \xi_{{\bf k}} \pm \sqrt{t_{h}^{2} + |{\bf g_{k}}|^{2}}
$
which are with twofold degeneracy. For a system with degenerate Fermi surfaces, the above FSTI cannot work and some perturbations should be introduced to lift the degeneracy while preserving the PHS and TRI symmetry of the system. To lift the degeneracy, we can introduce an imbalance in chemical potential $h = (\mu_{B}-\mu_{A})/2$ of the two layers with $\mu = (\mu_{A}+\mu_{B})/2$, then the single particle Hamiltonian changes into $h_{{\bf k}}' = h_{{\bf k}} + h \tau_{z}$ with Pauli matrix $\tau_{z}$ act on the layer part, hence the four eigenvalues of the imbalanced Hamiltonian $E_{i}({\bf k})$ ($E_{1}({\bf k}) < E_{2}({\bf k}) < E_{3}({\bf k}) < E_{4}({\bf k})$) have no degeneracy and the FSTI can work well. The sign of the pairing gap on the Fermi surfaces is ${\rm sgn}(\delta_{1,3})=-1$, ${\rm sgn}( \delta_{2,4})=1$ for $|{\bf g_{k}}| > h$ and ${\rm sgn}(\delta_{1,2})=-1$, ${\rm sgn}(\delta_{3,4})=1$ for $|{\bf g_{k}}| < h$ (More details are included in the APPENDIX).

\begin{figure}
\includegraphics[width=6.5cm, height=6.5cm]{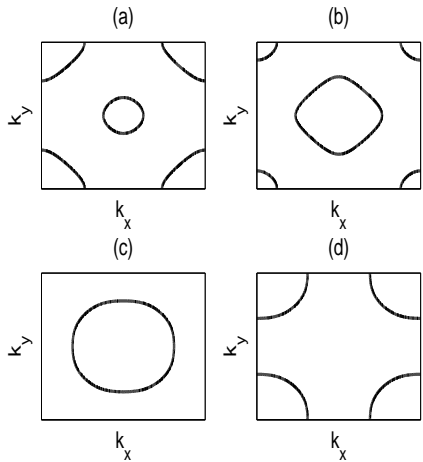}
\caption{Fermi surfaces of $E_{1,3}({\bf k})$ in the first Brillouin Zone with $h=0$ for $t_{h}=2t$, $\lambda= t$ with (a) $\mu = -t$, (b) $\mu=t$, (c) $\mu=-4t$ and (d) $\mu=4t$.  The number of TRI points enclosed by Fermi surfaces is two for (a), (b) and one for (c),(d) respectively.} \label{fig.1}
\end{figure}

For $|{\bf g_{k}}| < h$, the imbalanced Hamiltonian cannot adiabatically changes into the balanced case by taking the limit $h \rightarrow 0$, this indicates that there exists topological phase transitions as $h$ increasing. In this work, we only concentrate on the balanced case and leave the imbalance case for future discussion.

For $|{\bf g_{k}}| > h$, the imbalanced Hamiltonian can be adiabatically changed into the balanced case by taking the limit $h \rightarrow 0$ without changing any topological properties, thus the sign of the pairing gap on the Fermi surfaces for the balanced case is the same as the imbalanced case. Such that the FSTI only depends on the number of the TRI points enclosed by the Fermi surfaces of the $E_{1,3}({\bf k})$. Fig.\ref{fig.1} shows the Fermi surfaces of $E_{1,3}({\bf k})$  for various cases. The number of the TRI points enclosed by the Fermi surfaces is $2$ for $(a), (b)$ and $1$ for $(c),(d)$. According to analysis of the relationship between $E_{1}({\bf k})$ and $E_{3}({\bf k})$, we have $m_{1} + m_{3}$ amounts to $0$ for $t_{h}> 4t + |\mu|$ , $2$ for $t_{h} < 4t - |\mu|$ and $1$ for $4t - |\mu| < t_{h} < 4t + |\mu|$. Therefore, the superfluid phases are topological nontrivial when the interlayer hopping satisfies the topological condition $4t - |\mu| < t_{h} < 4t + |\mu|$ in the weakly pairing limit.

\section{Topological Superfluid Phase}
\begin{figure}
\includegraphics[width=8.5cm, height=4cm]{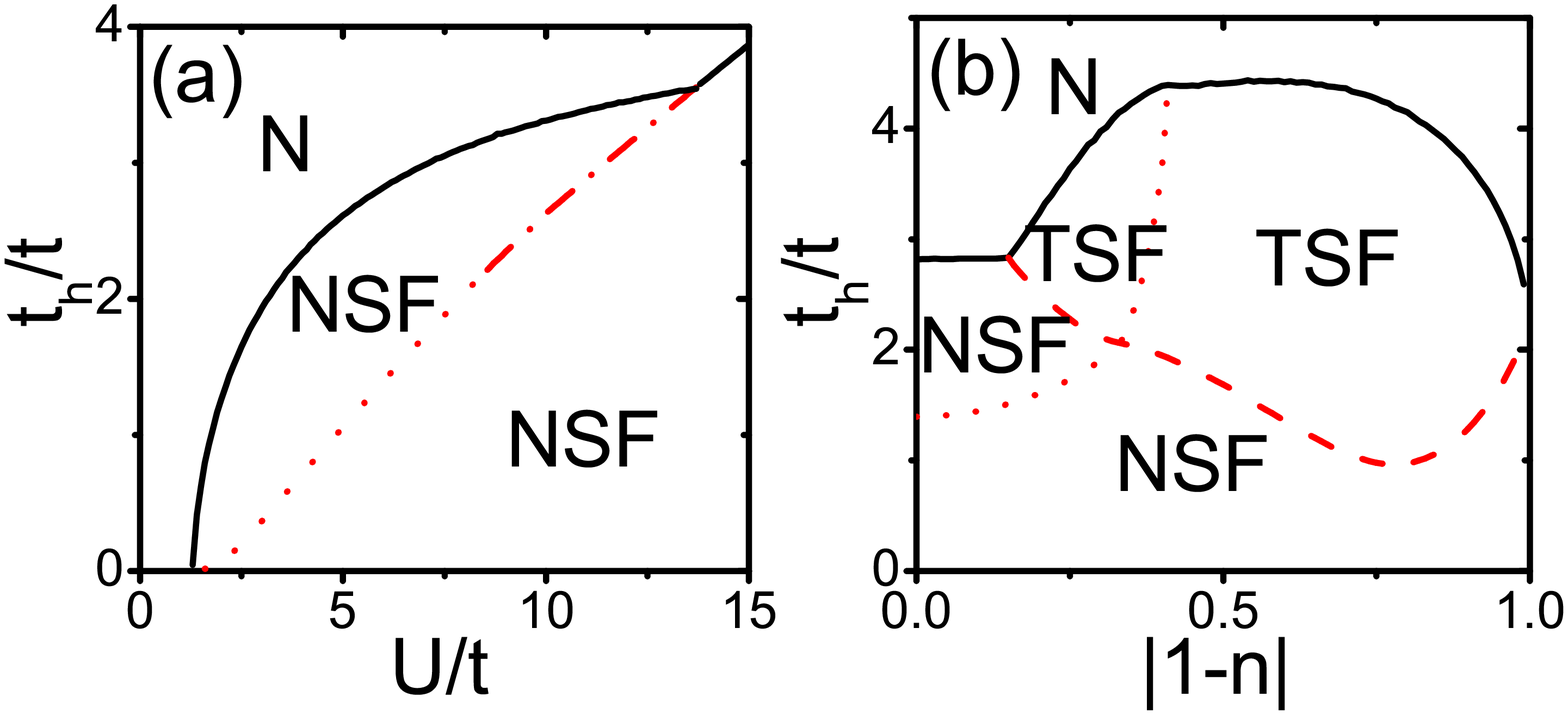}
\caption{(left)The phase diagram as a function of $U$ and $t_{h}$ with $\lambda = t$ and $n=1$. (right)The phase diagram as a function of filling fraction $n$ and the interlayer hopping $t_{h}$ with $\lambda = t$ and $U=6 t$. The dotted curve represents the gapless boundary and the dashed-dotted curve represents first-order phase boundary. The dashed curve represents topological phase boundaries. The superfluid phases are all topological nontrivial for $n=1$. There are two topological superfluid regions separated by a dotted curve when the filling fraction deviates from the half filling.} \label{fig.2}
\end{figure}

\begin{figure}
\includegraphics[width=8.5cm, height=6.5cm]{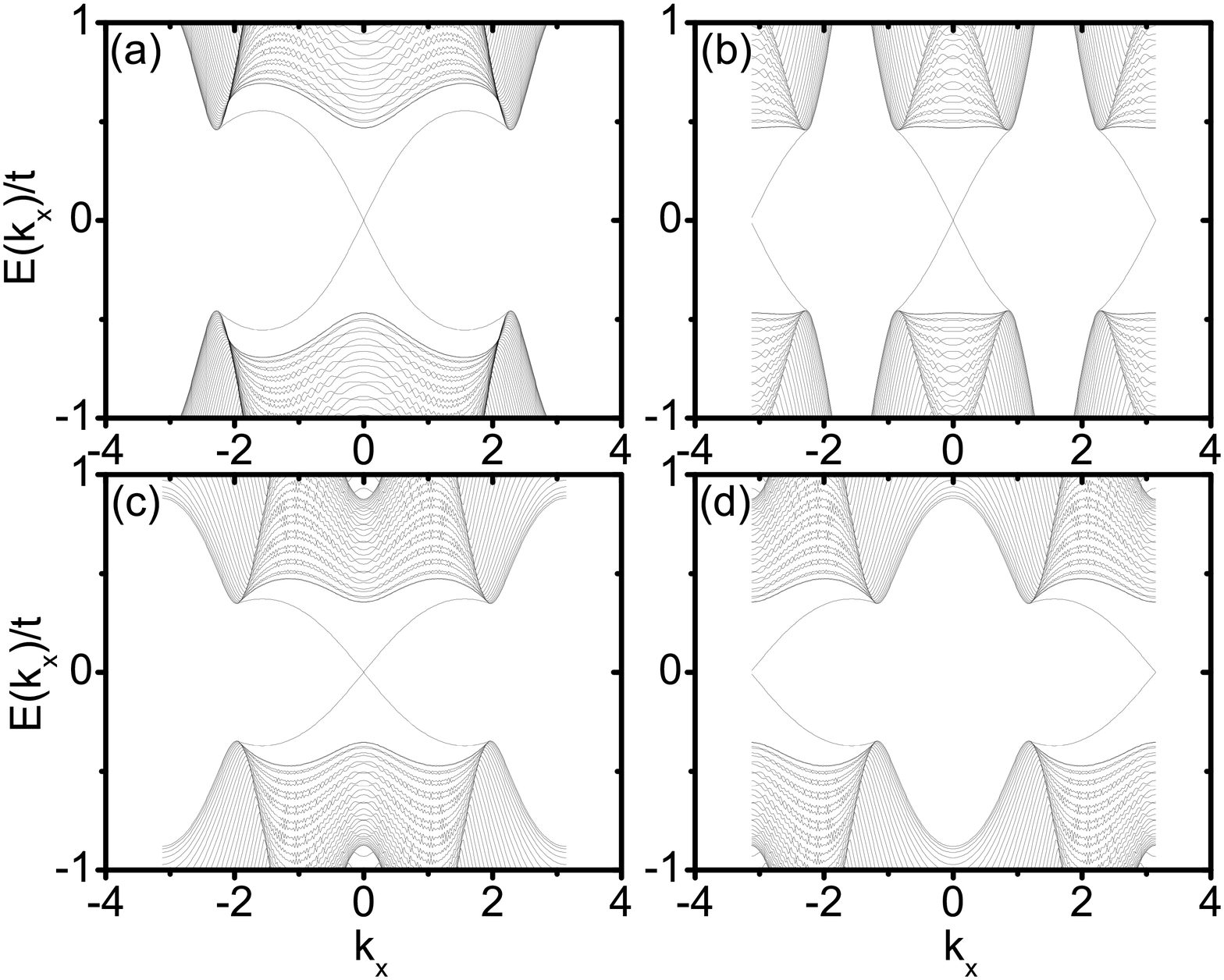}
\caption{The spectrum of Hamiltonian \ref{1} in a strip geometry for (a) $\mu=-4 t$, $t_{h}=3 t$ and (b) $\mu=0$, $t_{h}= 3 t$ (c) $\mu=4 t$, $t_{h}=5 t$ (d) $\mu=-4 t$, $t_{h}=5 t$ with $\lambda = t$ and $\Delta = t$.} \label{fig.3}
\end{figure}

The topological phase transition only occurs at the gapless points with the changing of the topological invariants. Thus, the gapless conditions show the significance of determining the topological invariants of the superfluid phases in the phase diagram. Since the superfluid phases are topological trivial in the absence of interlayer hopping ($t_{h}=0$), the topological invariant of the superfluid phase near the boundaries between the normal and superfluid phases can be determined by the FSTI. Therefore, the topological invariants of all the superfluid phases in the phase diagram can be determined.

First, we show the phase diagram as a function of on-site interaction ($U$) and interlayer hopping ($t_{h}$) for half filling case in Fig.\ref{fig.2}(a). The dotted curve represents the gapless boundary and the dashed-dotted curve represents first-order phase boundary along which the thermodynamic potential has two degenerate minima. By analyzing the relationship between $\mu$ and $t_{h}$ along the boundary between the normal and the superfluid phases, we have $N_{2D} = 1$ for all the superfluid phase which indicates that there is no topological nontrivial superfluid phase for the half filling case.

Second, in order to find the topological nontrivial superfluid phase, we investigate the case as deviating from the half filling. As the system has particle-hole symmetry at half filling, the phase diagram has an axial symmetry. Fig.\ref{fig.2}(b) shows the phase diagram as a function of filling fraction ($n$) and the interlayer hopping ($t_{h}$). The dashed and the dotted curves represent the gapless boundaries along which the gapless points are $(0,0)$ ($(\pi,\pi)$) and $[(0,\pi),(\pi,0)]$ for $\mu<0$ ($\mu>0$) respectively. Therefore, the chemical potential and the interlayer hopping satisfy the topological condition $4t - |\mu| < t_{h} < 4t + |\mu|$  at the superfluid side of the normal-superfluid boundary which is above the dashed curve. Without the interlayer hopping, the system can be considered as two independent SO coupled Fermi gases in which the superfluid phases are all topological trivial. As the interlayer hopping increases, there exists a topological phase transition when the filling fraction deviates from the half filling. As shown in Fig.\ref{fig.2}(b), there are two TSF phases. The two TSF phases are helical superfluid phases and have no difference in topology. The only difference between the two helical superfluid phases is that the TRI points enclosed by the Fermi surfaces are different in the weakly pairing limit.

As mentioned before, the TSF phases are also characterized by the gapless boundary modes. For a 2D TRI TSF phase, the boundary modes are gapless helical Majorana edge states, where time-reversal partners counterpropagate. Accordingly, each TRI topological defect of the TSFs carries a Kramers pair of Majorana fermions. To investigate the helical Majorana edge states, we calculate the spectrum of the Hamiltonian (\ref{1}) in a strip geometry where the edges are along the $y$ direction. Due to breaking of translation symmetry along the $y$ direction, $k_{y}$ is not a good quantum number while $k_{x}$ is a good quantum number, and then we perform a Fourier transformation along the $x$ direction only. Fig.\ref{fig.3} show the spectrum of the Hamiltonian as a function of $k_{x}$ for various $\mu$ and $t_{h}$ with $\lambda= t$ and $\Delta=t$.  Fig.\ref{fig.3}(a) and (d) are the cases of the $0<n<1$ while the Fig.\ref{fig.3}(c) is the case $1<n<2$. The three phases have a full gap in the bulk and gapless helical Majorana edge states counterpropagating at the boundary. Fig.\ref{fig.3}(b) is the  half filling case and there are two pairs of gapless Majorana edge states at the boundary, which indicates that the phase is topological trivial.

At last, we consider the effects of a weakly harmonic trapping potential under LDA. In the presence of harmonic trapping potential, the global chemical potential ($\mu_{{\bf i}}=\mu- m \omega ^{2} r_{{\bf i}}^{2}$) and the pairing gap are the functions of spatial coordinates. The topological invariants of the superfluid phases depend on the spatial coordinates. Therefore, the helical superfluid phase and topological trivial phases (normal phase and topological trivial superfluid phase) can coexist in the trapped region and a shell structure of topological phase separation emerges\cite{ J zhou, X yang1, X yang2}. The helical Majorana edge states, which can be called helical Majorana liquid, exists at the interfaces of the helical superfluid phase and topological trivial phases. The position of the helical Majorana liquid is controlled by the parameters of the system. The existence of the helical Majorana liquid will lead to nontrivial physical consequences\cite{Nakosai}. In addition, we only consider the interlayer hopping and the intralayer on-site attractive interaction for simplicity. When the interlayer interaction is taken into account, the phase diagram will be enriched. Furthermore, the helical superfluid phase with unconventional pairing like the $d_{x^{2}-y^{2}}$ might be stabilized when the on-site interaction is replaced by the nearest-neighbor interaction.

\section{CONCLUSIONS}
To conclude, we have constructed the phase diagram for a bilayer spin-orbit coupled Fermi gas loaded in a square optical lattice. We have found that the TRI superfluid phases are topological nontrivial when the filling fraction deviates from the half filling. There are helical edge states at the boundary of the TRI topological superfluid phase. In addition, we also have discussed the effects of a harmonic trapping potential. There is a shell structure of topological phase separation phenomena in the trapped region. We expect our work can provide a possible route to realize the helical superfluid phase by ultracold fermionic system in the future.

\section{ACKNOWLEDGMENTS}
We are very grateful to Wei Yi and Zhong Wang for useful discussions. This work was partially supported by China Postdoctoral Science Foundation Funded Project (Grant No.2012M520147).

\section{APPENDIX: DERIVATION OF THE FSTI}
The imbalanced single particle Hamiltonian is
\begin{eqnarray}
h_{{\bf k}}'=\left(
  \begin{array}{cc}
    \xi_{{\bf k}} + h + {\bf g_{k}} \cdot {\bf \sigma} & t_{h}  \\
    t_{h} &  \xi_{{\bf k}} - h - {\bf g_{k}}\cdot {\bf \sigma}  \\
  \end{array}
\right).
\end{eqnarray}

The four nondegenerate eigenvalues and the corresponding eigenvectors of $h_{{\bf k}}'$ are
\begin{eqnarray}
E_{i}({\bf k}) = \xi_{{\bf k}} \pm \sqrt{t_{h}^{2} + (|{\bf g_{k}}| \pm h)^{2}},
\end{eqnarray}
\begin{eqnarray}
|1,{\bf k}> = \left(
  \begin{array}{c}
    (g_{k}^{x} + i  g_{k}^{y}) \frac{|{\bf g_{k}}| + h - \sqrt{t_{h}^{2} +(|{\bf g_{k}}| +h )^{2}}}{t_{h}} \\
    \frac{|{\bf g_{k}}| + h - \sqrt{t_{h}^{2} +(|{\bf g_{k}}| +h )^{2}}}{t_{h}}  \\
    (g_{k}^{x} + i  g_{k}^{y})\\
    1\\
  \end{array}
\right);\\
|2,{\bf k}> = \left(
  \begin{array}{c}
    (g_{k}^{x} + i  g_{k}^{y}) \frac{|{\bf g_{k}}| - h + \sqrt{t_{h}^{2} +(|{\bf g_{k}}| -h )^{2}}}{t_{h}} \\
    -\frac{|{\bf g_{k}}| - h + \sqrt{t_{h}^{2} +(|{\bf g_{k}}| - h )^{2}}}{t_{h}}  \\
    -(g_{k}^{x} + i  g_{k}^{y})\\
    1\\
  \end{array}
\right);\\
|3,{\bf k}> = \left(
  \begin{array}{c}
    (g_{k}^{x} + i  g_{k}^{y}) \frac{|{\bf g_{k}}| - h - \sqrt{t_{h}^{2} +(|{\bf g_{k}}| - h )^{2}}}{t_{h}} \\
    - \frac{|{\bf g_{k}}| - h + \sqrt{t_{h}^{2} +(|{\bf g_{k}}| - h )^{2}}}{t_{h}}  \\
    -(g_{k}^{x} + i  g_{k}^{y})\\
    1\\
  \end{array}
\right);\\
|4,{\bf k}> = \left(
  \begin{array}{c}
    (g_{k}^{x} + i  g_{k}^{y}) \frac{|{\bf g_{k}}| + h + \sqrt{t_{h}^{2} +(|{\bf g_{k}}| +h )^{2}}}{t_{h}} \\
    \frac{|{\bf g_{k}}| + h + \sqrt{t_{h}^{2} +(|{\bf g_{k}}| +h )^{2}}}{t_{h}}  \\
    (g_{k}^{x} + i  g_{k}^{y})\\
    1\\
  \end{array}
\right),
\end{eqnarray}
with $E_{1}({\bf k})<E_{2}({\bf k})<E_{3}({\bf k})<E_{4}({\bf k})$.

The time-reversal matrix and the pairing gap matrix are
\begin{eqnarray}
\mathcal{T} =\left(
  \begin{array}{cc}
    i \sigma_{y} & 0 \\
    0 &  i \sigma_{y} \\
  \end{array}
\right),
{\bf \Delta} =\left(
  \begin{array}{cc}
    i \Delta \sigma_{y} & 0 \\
    0 &  -i\Delta \sigma_{y} \\
  \end{array}
\right).
\end{eqnarray}

By inserting the pairing gap and the eigenvectors into $\delta_{i,{\bf k}}$, we have
\begin{eqnarray}
\delta_{i,{\bf k}} =
\left\{
  \begin{array}{c}
    4(|{\bf g_{k}}| + h) \frac{|{\bf g_{k}}| + h - \sqrt{t_{h}^{2} +(|{\bf g_{k}}| + h )^{2}}}{t_{h}^{2}}, ~~~~i=1\\
    4(|{\bf g_{k}}| - h) \frac{|{\bf g_{k}}| - h + \sqrt{t_{h}^{2} +(|{\bf g_{k}}| - h )^{2}}}{t_{h}^{2}}, ~~~~~~~~2\\
    4(|{\bf g_{k}}| - h) \frac{|{\bf g_{k}}| - h - \sqrt{t_{h}^{2} +(|{\bf g_{k}}| - h )^{2}}}{t_{h}^{2}}, ~~~~~~~~3\\
    4(|{\bf g_{k}}| + h) \frac{|{\bf g_{k}}| + h + \sqrt{t_{h}^{2} +(|{\bf g_{k}}| + h )^{2}}}{t_{h}^{2}}, ~~~~~~~~4\\
  \end{array}
\right.
\end{eqnarray}

The sign of the pairing gap on the Fermi surfaces is ${\rm sgn}(\delta_{1,3})=-1$, ${\rm sgn}(\delta_{2,4})=1$ for $|{\bf g_{k}}| > h$ and ${\rm sgn}(\delta_{1,2})=-1$, ${\rm sgn}(\delta_{3,4})=1$ for $|{\bf g_{k}}| < h$.


\begin{thebibliography}{99}
\bibitem{gapphase} The phase shift between the pairing gaps of the two layers can be protected by the sign of a weak interaction $U_{AB} a_{i,\uparrow}^{\dag} a_{i,\downarrow}^{\dag} b_{i,\downarrow} b_{i,\uparrow} + h.c.$. In mean field framework, this term contribute a constant $ 2 U_{AB} |\Delta_{A} \Delta_{B}| \cos(\varphi_{A}-\varphi_{B})/ U^{2}$ ($\Delta_{A}$($\Delta_{B}$) is the pairing gap of the layer A(B), $\varphi_{A,B}$ are the phases of the two pairing gaps respectively) to the thermodynamic potential at zero temperature. Therefore, the phase shift between the pairing gaps of the two layers is $\pi$ for $U_{AB}<0$ and $0$ for $U_{AB}>0$.
\end{thebibliography}
\end{document}